\documentclass[aps,prb,10pt,twocolumn,showpacs,amsmath,amssymb,superscriptaddress]{revtex4-2}
\usepackage{graphicx}
\usepackage{dcolumn}
\usepackage{bm}
\usepackage{hyperref}
\hypersetup{colorlinks,citecolor=blue, filecolor=blue, linkcolor=blue , urlcolor=blue}

\begin{document}

\title{Strong and selective magnon-phonon coupling in van der Waals antiferromagnet CoPS$_3$
}

\author{Dipankar Jana}
    \email{jana.d02@nus.edu.sg}
    \affiliation{Laboratoire National des Champs Magn\'etiques Intenses, LNCMI-EMFL, CNRS UPR3228,Univ. Grenoble Alpes, Univ. Toulouse, Univ. Toulouse 3, INSA-T, Grenoble and Toulouse, France}
    \affiliation{Institute for Functional Intelligent Materials, National University of Singapore, 117544, Singapore}

\author{Diana~Vaclavkova}
    \affiliation{Laboratoire National des Champs Magn\'etiques Intenses, LNCMI-EMFL, CNRS UPR3228,Univ. Grenoble Alpes, Univ. Toulouse, Univ. Toulouse 3, INSA-T, Grenoble and Toulouse, France}

\author{Rajesh~Kumar~Ulaganathan}
    \affiliation{Institute of Physics, Academia Sinica, Taipei 10617, Taiwan}

\author{Raman~Sankar}
    \affiliation{Institute of Physics, Academia Sinica, Taipei 10617, Taiwan}

\author{Milan~Orlita}
    \affiliation{Laboratoire National des Champs Magn\'etiques Intenses, LNCMI-EMFL, CNRS UPR3228,Univ. Grenoble Alpes, Univ. Toulouse, Univ. Toulouse 3, INSA-T, Grenoble and Toulouse, France}
    \affiliation{Institute of Physics, Charles University, Ke Karlovu 5, Prague, 121 16, Czech Republic}

\author{Clement~Faugeras}
\affiliation{Laboratoire National des Champs Magn\'etiques Intenses, LNCMI-EMFL, CNRS UPR3228,Univ. Grenoble Alpes, Univ. Toulouse, Univ. Toulouse 3, INSA-T, Grenoble and Toulouse, France}

\author{Maciej~Koperski}
    \affiliation{Institute for Functional Intelligent Materials, National University of Singapore, 117544, Singapore}
    \affiliation{Department of Materials Science and Engineering, National University of Singapore, 117575, Singapore}

\author{M. E. Zhitomirsky}
    \affiliation{Universit´e Grenoble Alpes, CEA, Grenoble INP, IRIG, Pheliqs, 38000 Grenoble, France}
    \affiliation{Institut Laue-Langevin, F-38042 Grenoble Cedex 9, France}
    
\author{Marek Potemski}
    \email{marek.potemski@lncmi.cnrs.fr}
    \affiliation{Laboratoire National des Champs Magn\'etiques Intenses, LNCMI-EMFL, CNRS UPR3228,Univ. Grenoble Alpes, Univ. Toulouse, Univ. Toulouse 3, INSA-T, Grenoble and Toulouse, France}
     \affiliation{CENTERA, CEZAMAT, Warsaw University of Technology, 02-822 Warsaw, Poland} 
    \affiliation{Institute of High Pressure Physics, PAS, 01-142 Warsaw, Poland}

\begin{abstract}

The Raman scattering response of the biaxial antiferromagnet CoPS$_3$ has been investigated as a function of both magnetic field and temperature. The peaks observed in the low-frequency spectral range (90–200~cm$^{-1}$) have been identified as hybrid magnon–phonon excitations. The energies of the bare magnon and phonon modes and the effective coupling strengths between different excitation pairs have been determined. The strong and selective magnon-phonon interaction largely accounts for the pronounced splitting of two phonon-like modes observed at 152~cm$^{-1}$ and 158~cm$^{-1}$ in the antiferromagnetic phase of CoPS$_3$. Based on the identification of bare magnon excitations and their magnetic-field dependence, we propose an updated set of parameters for the effective exchange (J$_{eff}$~=~9.9~meV) and biaxial magnetic anisotropy (D~=~4.3~meV and E~=~-0.7~meV) and advocate for an apparent anisotropic $g$-factor (g$_x$=g$_y$=2, g$_z$=4) in CoPS$_3$ antiferromagnet.

\end{abstract}

\maketitle

\section{Introduction}

The interplay between lattice vibrations (phonons) and spin wave excitations (magnons) in magnetic materials has attracted significant attention due to its fundamental importance and potential applications in fields such as spintronics, magnonics, and quantum information processing~\cite{bozhko2020magnon, wang2023magnon, delugas2023magnon, streib2019magnon, shin2018phonon, man2017direct, dai2000magnon, lyons2023acoustically, li2021advances, Holanda2018, simensen2019magnon}. Investigating magnon and phonon excitations—and their possible hybridization—is particularly compelling in van der Waals (vdW) magnets, especially antiferromagnets~\cite{wang2020magnon, lyons2023acoustically, li2020observation, Vaclavkova2021, Liu2021, Cui2023, mai2021, vaclavkova2020magnetoelastic,  luo2023evidence, Pawbake2022, Jana2023Fe}. These materials are of interest not only from a fundamental perspective, particularly in the strictly two-dimensional (2D) limit, but also from an applied standpoint. In magnetically anisotropic vdW antiferromagnets, magnons are gapped excitations (with finite energy at $k = 0$), and these magnon gaps often lie at sufficiently high energies near optical phonon modes~\cite{Liu2021, Cui2023, mai2021, vaclavkova2020magnetoelastic}. Such systems are especially relevant for terahertz (THz) technologies, potentially at micro- and nanoscale dimensions. This is particularly the case when these magnons couple to optically active phonon modes, enabling efficient manipulation and detection of spin-lattice dynamics and possibly tunability of the modes’ amplitude and/or frequency~\cite{ilyas2024terahertz, Liu2021, Pawbake2022}.
Coupling between $k = 0$ magnon and phonon modes, resulting in the formation of magnon-polarons, has already been demonstrated in iron-based vdW antiferromagnets such as FePS$_3$ and FePSe$_3$, where magnon and phonon excitations nearly coincide in energy~\cite{Liu2021, Cui2023, Vaclavkova2021, Pawbake2022, Jana2023Fe, luo2023evidence}. Crucially, the identification of this coupling relies on detecting characteristic mode repulsion when effectively tuning the $k = 0$ magnon-like mode via applied magnetic fields. Magnon and phonon modes are also expected to nearly coincide in the cobalt-based vdW antiferromagnet CoPS$_3$~\cite{Wildes2023}, making it another candidate for exploring magnon-phonon interactions.

\begin{figure}[bt]
	\includegraphics[width=8.4cm]{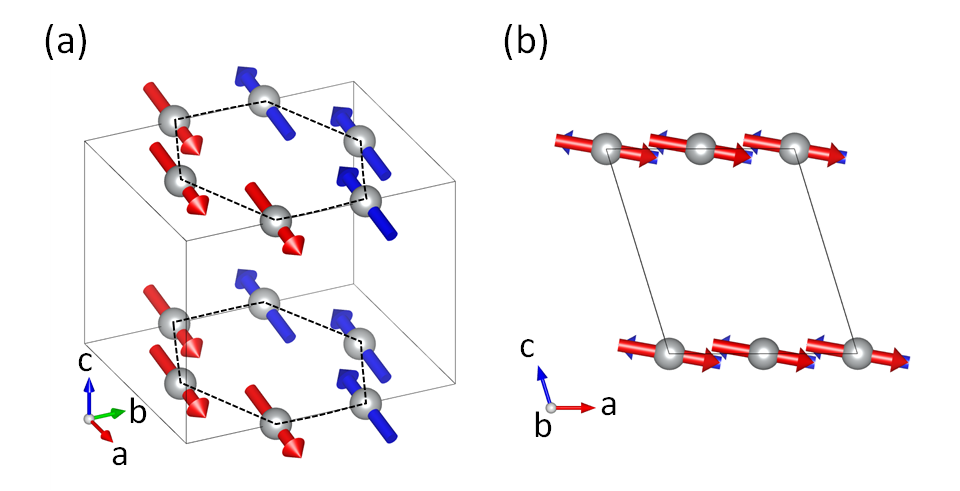}
	\caption{(a) The magnetic structure of CoPS$_3$ in its antiferromagnetic phase. Grey spheres, red, and blue arrows represent Co$^{2+}$ ions and the spin direction of two sublattices, respectively. The figure is created using the VESTA software package \cite{ Momma2011}. (b) View along the $b$-axis of CoPS$_3$ showing how Co$^{2+}$ spins are slightly canted out-of-plane~\cite{Wildes2023}.}
	\label{fig:fig1}
\end{figure}

Spin ordering and dynamics in CoPS$_3$ have been extensively investigated through magnetometry and a series of inelastic neutron scattering studies of this antiferromagnet~\cite{Wildes2017, Kim2020, Wildes2023, Wildes2025}. In the magnetically ordered phase, Co$^{2+}$ spins ($S=3/2$) are arranged on a honeycomb lattice, forming zigzag ferromagnetic chains along the $a$-axis. These chains are antiferromagnetically coupled along the $b$-axis (see Fig.~\ref{fig:fig1}). The magnetic moments are predominantly aligned along the crystallographic $a$-axis, although a slight out-of-plane canting of about 10° has also been suggested~\cite{Wildes2017}.
The theoretical modeling of the spin-wave dispersion has evolved over time~\cite{Wildes2017, Kim2020, Wildes2023, Wildes2025}. The most recent and comprehensive model is based on a two-dimensional (2D) spin-3/2 Hamiltonian that incorporates isotropic exchange interactions ($J_{ij}$) up to third-nearest neighbors in the $ab$-plane, as well as biaxial single-ion anisotropy described by two parameters, $D$ and $E$~\cite{Stevens1963, Wildes2025}:

\begin{equation}
\hat{\mathcal{H}} = \sum_{\langle ij \rangle} J_{ij} \mathbf{S}_i \cdot \mathbf{S}_j 
+ \sum_i \Bigl\{ D(S_{i}^z)^2\!+ E  \bigl[ (S_i^x)^2\! -\! (S_i^y)^2 ]\Bigr\} \,.
\label{H}
\end{equation}
Here, the $x$ and $y$ axes are defined along the crystallographic $a$ and $b$ directions, respectively, while $z$ is oriented perpendicular to the $ab$-plane (i.e., along the $c^*$-axis). This model reproduces well the spin-wave dispersion observed in CoPS$_3$ single crystals by setting a ferromagnetic first-neighbor exchange ($J_1$ = –1.37~meV), antiferromagnetic second-neighbor exchange ($J_2$ = 0.09~meV), and strong antiferromagnetic third-neighbor exchange ($J_3$ = 3.0~meV). The anisotropy parameters are determined to be $D = 6.07$ meV, $E = -0.77$ meV, favoring the in-plane spin alignment along the $x$-direction (i.e., the $a$-axis), consistent with $D > 0$ and $E < 0$. Notably, the magnon spectrum of CoPS$_3$ appears to be gapped (finite excitation energy at $k = 0$). Two, characteristic for a biaxial antiferromagnet, low energy magnon gaps have been estimated at 14~meV and 24~meV. While the 2D model effectively accounts for the inelastic neutron scattering data, it may still offer an incomplete description; nevertheless, recent estimates indicate that the interlayer exchange interactions are comparatively weak~\cite{Wildes2025}.

Bulk as well as few-layer and monolayers of CoPS$_3$ antiferromagnet have also been recently investigated with Raman scattering experiments focused on specific phonon modes~\cite{Liu2021Co}. The key reported observation was that a doubly degenerate phonon mode (at around 150~cm$^{-1}$) appearing due to the in-plane vibration of cobalt atoms is observed to split into two components when the temperature is lowered below the N\'eel temperature ($T_N = 120~K$). This effect, tentatively attributed to an anticipated crystal lattice modification induced by spin ordering has been used to trace the paramagnetic to antiferromagnetic phase transition in CoPS$_3$ specimens down to a monolayer.

Crucially, to date, neither inelastic neutron scattering nor Raman spectroscopy investigations of CoPS$_3$ have considered the potentially significant role of magnon–phonon hybridization in this antiferromagnet. This omission is particularly noteworthy given the analogies between CoPS$_3$ and Fe-based layered antiferromagnets (FePS$_3$ and FePSe$_3$). As in CoPS$_3$, the Fe-based antiferromagnets display magnon excitations near phonon modes, with clear evidence of magnon–phonon hybridization effects d~\cite{Liu2021, Cui2023, mai2021, vaclavkova2020magnetoelastic}.

In this paper, we report a Raman scattering study of low-energy excitations (in the spectral range of 90–200~cm$^{-1}$) in the van der Waals antiferromagnet CoPS$_3$. The measurements were performed as a function of the magnetic field, applied along different crystallographic directions, and as a function of temperature. In addition to the previously reported phonon-like excitations, we identify two additional Raman peaks consistent with zone-center ($k = 0$) magnon-type modes. All observed features are interpreted as arising from coupled magnon–phonon excitations. By analyzing their evolution with magnetic field and temperature, we extract characteristic coupling strengths and determine the energies of the uncoupled (bare) magnon and phonon modes. In particular, we find that the coupling involving a nearly degenerate phonon doublet, centered around 157~cm$^{-1}$, is exceptionally strong and selective: one component of the doublet hybridizes with the lower-energy magnon, while the other couples to the higher-energy magnon. Notably, the phonon doublet, nearly degenerate in the uncoupled case, becomes distinctly split as a result of the hybridization with magnon modes. Based on the extracted bare magnon-gap energies and their magnetic-field evolution, we propose an updated set of exchange interaction and magnetic anisotropy parameters for CoPS$_3$. Furthermore, our results point to a strongly anisotropic $g$-factor in this van der Waals antiferromagnet.

\section{Experimental details}

The CoPS$_3$ single crystals were grown by the chemical vapor transport method, using iodine as a transport agent. Initially, the polycrystalline powders were synthesized by a solid-state synthesis process under high vacuum conditions. The high-purity starting materials cobalt powder (99.999$\%$), phosphorus powder (99.999$\%$), and sulfur powder (99.999$\%$) were weighted at a stoichiometric ratio and sealed into the quartz tube with a diameter of 22~mm with 10$^{-3}$ Torr pressure. The mixed compounds were heated and grained twice at 400~$^{\circ}$C and 600~$^{\circ}$C to make a single-phase compound. The 200~mg iodine was added into the polycrystalline samples and sealed by the tube dimension of 20~mm~x~22~mm~x~400~mm with 10$^{-3}$ Torr. The tube was kept for growth at a two-zone furnace with a temperature range of 700~$^{\circ}$C and 600~$^{\circ}$C for 200~hrs. After completing the growth process, the temperature of the furnace was reduced to room temperature at a rate of 2~$^{\circ}$C/min. The quartz tube was broken inside the argon-filled glovebox, and the crystals were collected. A bulk crystal of $\approx$ 2~mm x 2~mm dimension was placed either onto a cold finger of a helium flow cryostat for temperature-dependent measurements or inside a liquid helium-cooled homemade setup for magneto-optical investigations where the resistive magnet can reach 30~T.

The micro-optical arrangements have been used for Raman scattering measurements as a function of temperature or a magnetic field in different configurations. A continuous wave semiconductor-based laser operating at a wavelength of $\lambda=515$~nm is focused on the sample with a microscope objective of numerical aperture of $=0.5$ (helium flow cryostat) or $=0.83$ (magneto-Raman measurements). The scattered signals were collected using the same objective, dispersed with a 0.7~m long monochromator with 2000~l/mm grating, and detected with a nitrogen-cooled charge-coupled device camera. A set of reflection-based Bragg filters is used in both the excitation and collection paths to clean the laser line and reject the backscattered laser. For in-plane magnetic field-dependent Raman scattering measurements, the sample was mounted in such a way that the magnetic field is parallel or perpendicular (or at some angle) to a particular edge of the crystal, with a presumption that the edge corresponds to a certain crystal axis.

\section{Experimental results and discussion}

\begin{figure*}[htp]
\centering
\includegraphics[width=16.8cm]{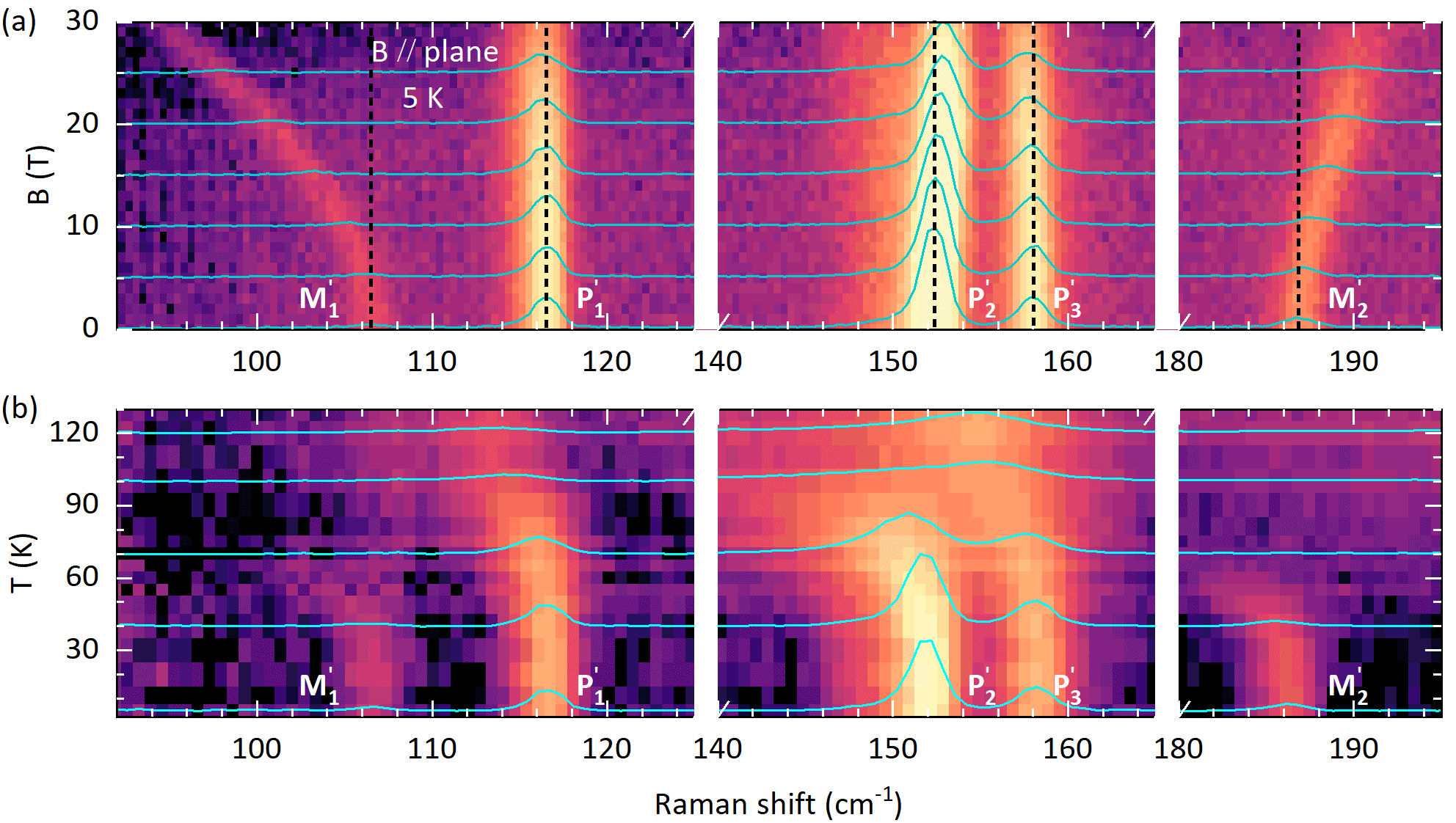}
	\caption{ False color map of the Raman scattering response of CoPS$_3$ antiferromagnet together with selected characteristic spectra. The coupled magnon-like modes are labeled as M$^{'}_i$ while the phonon-like modes are labeled as P$^{'}_i$. (a) Low temperature (5~K) data collected as a function of the magnetic field (B), applied plausibly along the in-plane crystal a-axis. Vertical dashed lines are drawn at the central energies of the phonon-like features measured at zero magnetic field. This is to emphasize the sensitivity, although weak, of the P$^{'}_i$ modes to the applied magnetic field. Common red or blue B-shifts for different pairs of P$^{'}_i$ and M$^{'}_i$ modes, point out the repulsion and thus selective coupling of the M$^{'}_1$ with P$^{'}_1$ and P$^{'}_3$ but M$^{'}_2$ with P$^{'}_2$. (b) Evolution of M$^{'}_i$ and P$^{'}_i$ modes with temperature (at $B=0~T$). Red shifts and softening of M$^{'}_i$ modes underline their magnon-like character.}     
	\label{fig:fig2}
\end{figure*}

Magnons (spin waves) and phonons are two characteristic low-energy excitations in antiferromagnets that, in the limit of zero wavevectors ($k=0$), can be effectively traced using Raman scattering experiments, especially when external parameters such as temperature or magnetic field are varied.

Our Raman scattering experiments on CoPS$_3$ crystals mainly focus on excitations at relatively low energies, in the spectral range of $90 - 200$~cm$^{-1}$. Two representative datasets are illustrated in Fig.~\ref{fig:fig2}. Spectra measured at base temperature ($T=5$~K), as a function of the applied magnetic field, B~=~$0$~-~$30$~T, are shown in Fig.~\ref{fig:fig2}a, whereas the spectral evolution with temperature, T~$= 5$~-~$130$~K, in the absence of the magnetic field, is presented in Fig.~\ref{fig:fig2}b (See Fig.~S1a of SM~\cite{SuppInfo} for Raman scattering over a broad spectral range). As mentioned above, we have carried out the magneto-Raman scattering experiments in various configurations of the field directions relative to the sample edges; the spectral evolution with the magnetic field changes for different configurations (See Fig.~S2 of SM~\cite{SuppInfo}). The data presented in Fig.~\ref{fig:fig2}a has been specifically selected for its most pronounced spectral changes in response to the magnetic field. As discussed below, this dataset corresponds most closely to the configuration of the magnetic field applied in the plane, along the crystal \textit{a}-axis, i.e., the spin alignment direction.

Focusing on the basic spectra (at T = $5$~K, B = $0$~T) we first recognize the characteristic Raman peaks, marked here as P$^{'}_1$, P$^{'}_2$, and P$^{'}_3$, which have been previously observed and assigned to phonon-type excitations~\cite{Liu2021Co}. Overlooked in previous Raman scattering studies of CoPS$_3$ are two other, weaker intensity peaks, M$^{'}_1$ and M$^{'}_2$, which are, correspondingly, centered at energies $\omega_{M_1}^{'}$=106~cm$^{-1}$ and $\omega_{M_2}^{'}$=186~cm$^{-1}$. As shown in Fig.~\ref{fig:fig2}a, the M$^{'}_1$ and M$^{'}_2$ peaks display pronounced energy shifts with the applied magnetic field: M$^{'}_1$ undergoes a redshift, while M$^{'}_2$ shows a blueshift. Moreover, the M$^{'}_1$ and M$^{'}_2$ resonances are considerably softened upon increasing temperature and are hardly observed at temperatures above $\approx60$~K (see Fig.~\ref{fig:fig2}b). These characteristic behaviors point towards the magnon origin of the M$^{'}_1$ and M$^{'}_2$ modes. Notably, the energies of these modes  (at T = $5$~K, B = $0$~T) are in reasonable agreement with those reported for the two low-energy magnon gap excitations observed in recent neutron scattering experiments~\cite{Wildes2023}.

\subsection{Magnon-phonon coupling}

\begin{figure}[t]
\centering
\includegraphics[width=8.4cm]{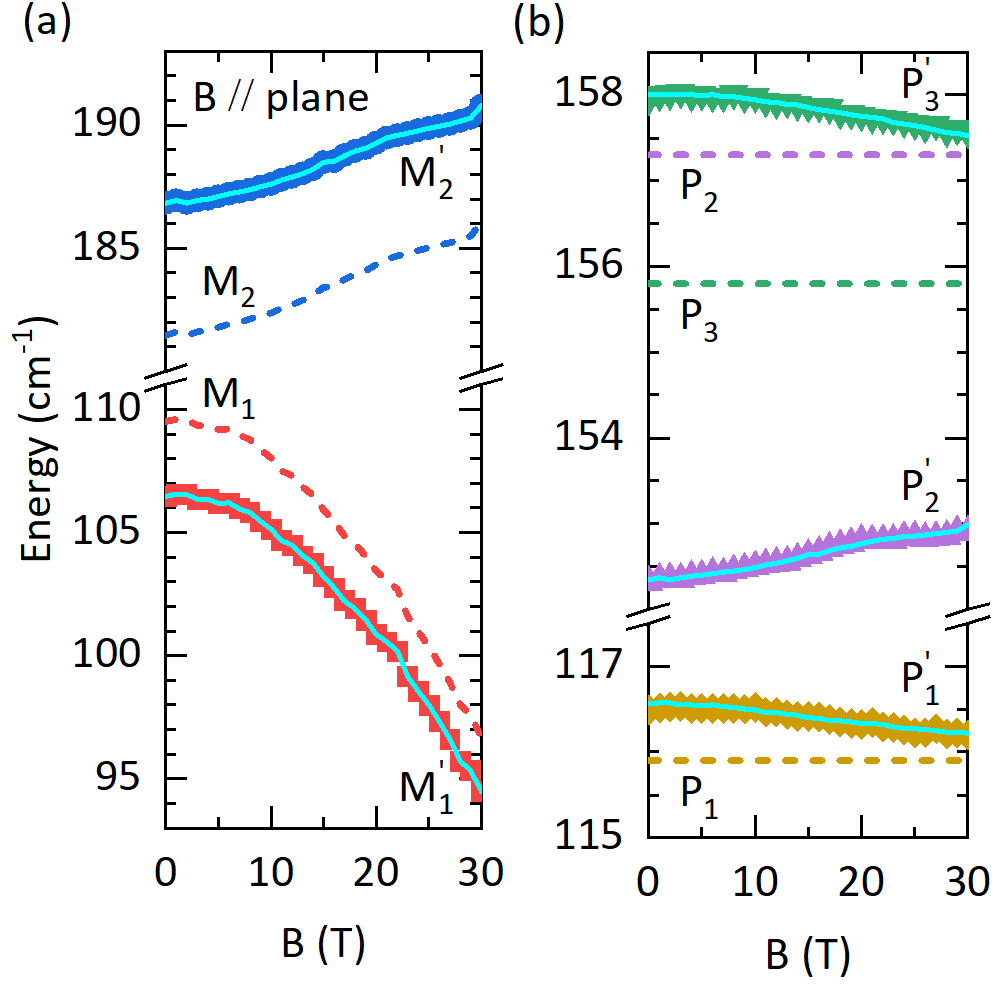}
\caption{Peak energies of (a) the magnon-like modes (M$^{'}_i$) and (b) the phonon-like modes (P$^{'}_i$) as a function of the magnetic field applied along the plane are shown by solid symbols. The solid and dashed lines display the simulated magnetic field dependence of coupled and bare mode energies, respectively, as calculated using Eq.~(\ref{Ham}).}
\label{fig:Fig3}
\end{figure}

In the following, we demonstrate that the observed Raman scattering modes are \textit{de facto}, not pure phonon or magnon excitations but represent the coupled magnon-phonon modes: P$^{'}_1$, P$^{'}_2$, and P$^{'}_3$ should be seen as phonon-like whereas M$^{'}_1$ and M$^{'}_2$ as magnon-like modes. The overall coupling scheme can already be deduced from the raw data shown in Fig.~\ref{fig:fig2}a. Closer inspection of the data indicates that the application of the magnetic field affects not only the M$^{'}_1$ and M$^{'}_2$ modes but also leads to shifts, although weaker, of P$^{'}_1$, P$^{'}_2$, and P$^{'}_3$ resonances. This is emphasized in Fig.~\ref{fig:Fig3} in which the energy positions of the observed Raman peaks are plotted as a function of the magnetic field. Markedly, the P$^{'}_2$ mode displays an upward shift with the magnetic field as the M$^{'}_2$ mode does, whereas the P$^{'}_1$, and P$^{'}_3$ modes experience downward shifts, following the field dependence of the M$^{'}_1$ mode. These observations direct us to the proposal of the selective magnon-phonon coupling in CoPS$_3$ antiferromagnet: the P$^{'}_2$ and M$^{'}_2$ modes result from the coupling of the bare P$_2$ and M$_2$ modes, whereas the bare P$_1$ and P$_3$ modes couple to the bare M$_1$ mode. In line with the earlier approach~\cite{Cui2023, Jana2023Fe}, used to describe the selective magnon-phonon coupling in FePSe$_3$ antiferromagnet, the following Hamiltonian is applied to reproduce the apparent energies of our coupled modes:
\begin{equation}
H_{5\times5} = \begin{bmatrix}
\omega_{M_1}& \delta_1 & \delta_2 & 0 & 0 \\
\delta_1 &\omega_{P_1} & 0 & 0 & 0 \\
\delta_2 & 0 & \omega_{P_3} & 0 & 0 \\
0 & 0 & 0 & \omega_{M_2} & \delta_3 \\
0 & 0 & 0 & \delta_3 & \omega_{P_2} \\
\end{bmatrix}
\label{Ham}
\end{equation}

Here, $\omega_{M_1}$ and $\omega_{M_1}$ correspond to the energy of the bare magnon modes M$_1$ and M$_2$ respectively. $\omega_{P_1}$, $\omega_{P_2}$, and $\omega_{P_3}$ correspond to energy of the bare phonon modes P$_1$, P$_2$, and P$_3$ respectively. $\delta_1$, $\delta_2$, and $\delta_3$ are the coupling parameters that control the hybridization of the M$_1$ magnon mode with the P$_1$ phonon mode, the M$_1$ magnon with the P$_3$ phonon, and the M$_2$ magnon with the P$_2$ phonon, respectively. Using this Hamiltonian, the eigenstates of the coupled magnon-phonon modes can be derived and eventually matched to those observed in the experiment. Such a procedure has been applied to the entire ensemble of the data points, at different magnetic fields, shown in Fig.~\ref{fig:Fig3} (See section~S3 of SM~\cite{SuppInfo}). To reproduce this set of data points, we assumed that the parameters $\delta_1$, $\delta_2$, $\delta_3$, $\omega_{P_1}$, $\omega_{P_2}$, and $\omega_{P_3}$ are independent of the magnetic field while allowing the energies, $\omega_{M_1}$ and $\omega_{M_2}$, of the bare magnon modes to vary with the field strength. The resulting data simulation is shown in Fig.~\ref{fig:Fig3}. The measured energies of coupled modes are best reproduced when fixing $\delta_1=2.5$~cm$^{-1}$, $\delta_2=10.8$~cm$^{-1}$, $\delta_3=12.5$~cm$^{-1}$, $\omega_{P_1}=116$~cm$^{-1}$, $\omega_{P_2}=157$~cm$^{-1}$, and $\omega_{P_3}=156$~cm$^{-1}$, and the $\omega_{M_1}$ and $\omega_{M_2}$ dependences as shown in Fig.~\ref{fig:Fig3}a. At zero magnetic field, the energies of bare magnon modes are found to be $\omega_{M_1}~(B=0~T)= 109.5$~cm$^{-1}$ and $\omega_{M_2}~(B=0~T)= 182$~cm$^{-1}$.

\begin{figure}[t]
\centering
\includegraphics[width=8.4cm]{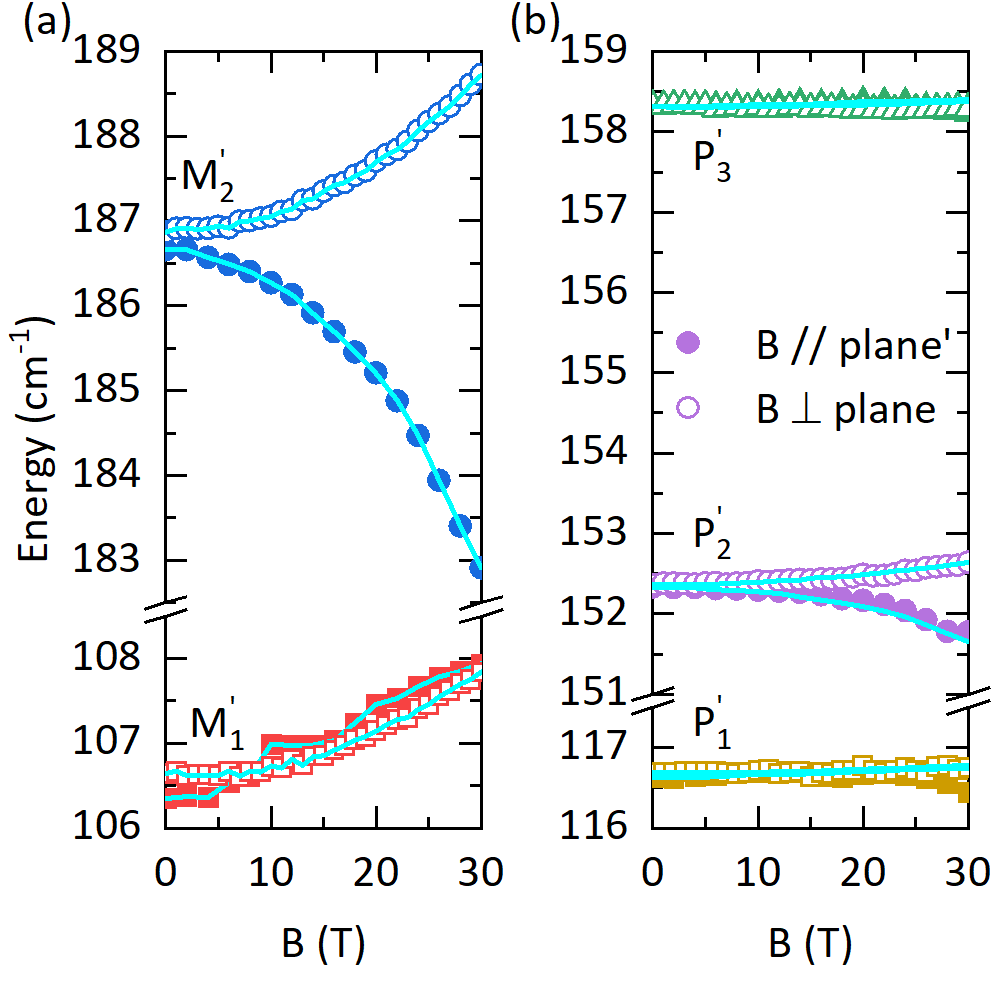}
\caption{Peak energies of (a) the magnon-like modes (M$^{'}_i$) and (b) the phonon-like modes (P$^{'}_i$) as a function of the magnetic field applied along a different in-plane direction (solid symbols) and out-of-plane direction (open symbols). The corresponding false color maps are presented in Fig.~S2d and Fig.~S3a, respectively. Solid lines display the simulated peak energy of the coupled modes as calculated by solving Eq.~(\ref{Ham}).}
\label{fig:Fig4}
\end{figure}

Moreover, when setting the above values for $\delta_i$, $\omega_{P_i}$, and $\omega_{M_i}$~($B=0~T$) parameters we reproduce the evolution of the coupled modes measured as a function of the magnetic field applied along a different in-plane direction and along the out-of-plane direction (see Fig.~\ref{fig:Fig4}). Notably, the extracted $\omega_{M_1}$(B) and $\omega_{M_2}$(B) dependences in these latter configurations differ from those obtained for the data shown in Fig.~\ref{fig:Fig3}a. 
The configuration-dependent B-evolutions of the extracted bare magnon modes are discussed in the next section. At this point, we highlight the role of magnon-phonon coupling on the apparent positions of the phonon-like modes, of the P$^{'}_2$, and P$^{'}_3$ modes. Whereas the observed energy positions of these modes are $\omega_{P'_2}=152$~cm$^{-1}$ and $\omega_{P'_3}=158$~cm$^{-1}$ their bare counterparts are almost degenerate, at $\omega_{P_2}=157$~cm$^{-1}$ and $\omega_{P_3}=156$~cm$^{-1}$, respectively. We, therefore, conclude that the observed separation
($\Delta=6$~cm$^{-1}$ between the P$^{'}_2$, and P$^{'}_3$ modes is at large the result of the magnon-phonon coupling. It is a consequence of the remarkably strong coupling parameters $\delta_2=10.8$~cm$^{-1}$ and $\delta_3=12.5$~cm$^{-1}$, even though the hybridizing bare magnon and phonon modes are quite separated in energy. Interestingly, our data modelling implies that the P$^{'}_2$ mode is located $6$~cm$^{-1}$ below the P$^{'}_3$ mode whereas the energy ordering of their bare counterparts, P$_2$ and P$_3$ separated by $1$~cm$^{-1}$, is inverted. The observed separation between the P$^{'}_2$, and P$^{'}_3$ modes measured in CoPS$_3$ at low temperatures has been previously attributed to the effect of magnetostriction, possibly associated with the spin ordering phase. As discussed later, the magnetostriction can be at the origin of the $1$~cm$^{-1}$ splitting, which we conclude for the bare P$_2$ and P$_3$ modes. Still, the observed $6$~cm$^{-1}$ separation between the apparent P$^{'}_2$ and P$^{'}_3$ modes is mainly due to magnon-phonon interaction.

\begin{figure*}[tb]
\centering
\includegraphics[width=16.8cm]{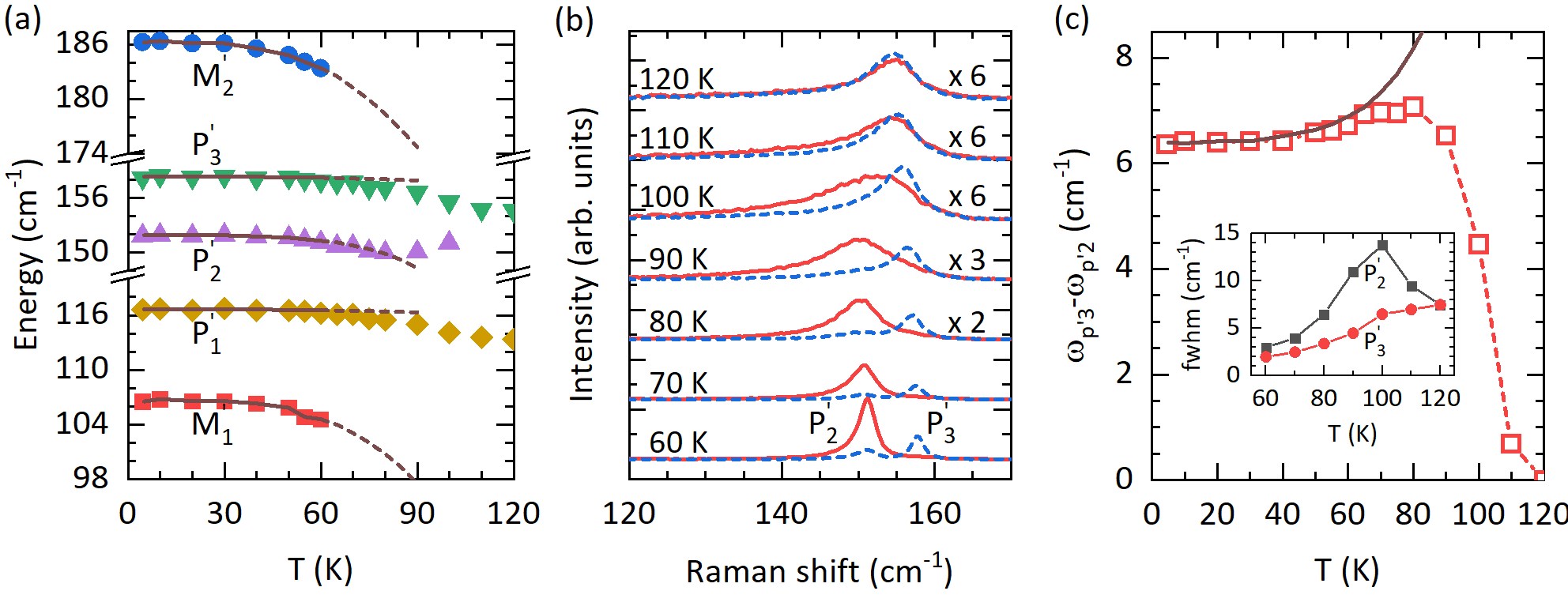}
\caption{(a) Peak energies of coupled modes as a function of temperature (solid symbols). The solid lines correspond to the simulated peak energies of the coupled modes calculated using Eq.~(\ref{Ham}). The bare magnon modes follow -T$^{3.5}$ dependence as shown in Fig.~S5 of the SM~\cite{SuppInfo}. The extrapolated temperature dependence of the bare magnon modes is used to estimate the coupled mode energies at elevated temperatures, as shown by the dashed line. (b) Linear polarization-resolved Raman scattering spectra highlighting the evolution of P$^{'}_2$ and P$^{'}_3$ modes as a function of temperature below T$_N$. (c) The energy difference between the P$^{'}_2$ and P$^{'}_3$ modes as a function of temperature. The solid line corresponds to the energy difference obtained from the simulation. The width (fwhm) of P$^{'}_2$ and P$^{'}_3$ modes as a function of temperature is shown in the inset.}
\label{fig:Fig5}
\end{figure*}

Further insights into the low-energy excitations in CoPS$_3$ are gained by analyzing the Raman scattering spectra measured at different temperatures. Peak positions (energies at maxima of peak intensities) of Raman scattering peaks extracted from the spectra illustrated in Fig.~\ref{fig:fig2}b are shown in Fig.~\ref{fig:Fig5}a. First, we focus on the temperature range T=5~K-~60~K where all P$^{'}_i$ and M$^{'}_i$ peaks are visible in the spectra and simulate their energy positions following the proposed model of magnon-phonon coupling. We assume that the bare phonon modes and coupling coefficients are temperature-independent in this temperature range and use the same set of parameters, as specified above. As shown with solid lines in Fig.~\ref{fig:Fig5}a, the experimental data are reasonably well reproduced and we find that the energies of bare magnon modes exhibit softening with temperature according to the often encountered rule~\cite{kobler2020}: $\omega_{M_i}~(T)~-~\omega_{M_i}~(T=0~K)\approx-~T^{3.5}$ (see Fig.~S5 of SM~\cite{SuppInfo}).

These magnon mode energies drop to zero at N\'eel temperature, which implies that before reaching T$_N$=120~K the M$_2$ mode crosses all phonon modes. The temperature evolution of the phonon-like modes observed above T$=60$~K cannot be solely explained by our current model of magnon-phonon coupling, particularly in its assumption of temperature-independent bare phonon modes and coupling coefficients. The signatures of magnon-phonon coupling are, however, clearly visible, especially when focusing on the P$^{'}_2$ and P$^{'}_3$ resonances. This can be conveniently demonstrated when inspecting the P$^{'}_2$ and P$^{'}_3$ peaks in the polarization-resolved measurements (see Fig.~\ref{fig:Fig5}b). The Raman scattering spectra of CoPS$_3$ are known to follow the characteristic, linear-polarization selection rules~\cite{Liu2021Co} (see also Fig.~S1b of SM~\cite{SuppInfo}). In particular, the P$^{'}_1$, P$^{'}_2$, and M$^{'}_2$ modes are observed in the same linear polarization of the excitation laser, whereas the P$^{'}_3$ and M$^{'}_1$ mode appears in the orthogonal polarization. Peak positions (energies corresponding to the maxima of peak intensities) and peak widths (full width at half maxima, fwhm) are shown in Fig.~\ref{fig:Fig5}b. The temperature evolution of the energy separation ($\omega_{P'_3}$-$\omega_{P'_2}$) between the P$^{'}_2$ and P$^{'}_3$ peaks is plotted in Fig.~\ref{fig:Fig5}c while the FWHM is shown in the inset. As can be seen in Fig.~\ref{fig:Fig5}b, at temperatures up to $80$~K both P$^{'}_2$ and P${'}_3$ peaks display redshifts with temperature. The shift of the P$^{'}_2$ peak (induced by the M$_2$ magnon closely approaching the P$_2$ phonon) is, however, more pronounced than the shift of the P${'}_3$ mode (which results from the increasing separation between the P$_3$ and M$_1$ modes). 
This explains the initial increase of ($\omega_{P'_3}$-$\omega_{P'_2}$) from 6~cm$^{-1}$ (at 5~K) to 7~cm$^{-1}$ (at 80~K) as illustrated in Fig.~\ref{fig:Fig5}c. The strong broadening of the P$^{'}_2$ peak, illustrated in the inset of Fig.~\ref{fig:Fig5}c, is observed at temperatures around 100~K, when the bare M$_2$ and P$_2$ modes are expected to overlap. This is another consequence of the magnon-phonon coupling in CoPS$_3$. When temperature rises above 80~K, the separation between the P$^{'}_3$ and P$^{'}_3$ modes sharply decreases and drops to zero at 120~K. Shrinkage of ($\omega_{P'_3}$-$\omega_{P'_2}$) when approaching the paramagnetic phase, and thus losing the magnon resonances and the magnon-phonon coupling strength, is expected within our model of magnon-phonon hybridization. This model, however, does not explain the absence of the splitting between the P$_2$ and P$_3$ modes at T=120~K as well as the actual energy position of the phonon modes observed at 120~K. Otherwise, they would be expected to correspond to the bare phonon modes, and in particular, we would expect to observe the 1~cm$^{-1}$ splitting between the P$_2$ and P$_3$ peaks in the paramagnetic phase. Apart from the strong magnon-phonon coupling, other mechanisms, e.g., magnetostriction, likely exist, which can also create the 1~cm$^{-1}$ splitting of the phonon resonances in the antiferromagnetic phase of CoPS$_3$.

 \subsection{Bare magnon modes}

Having deconvoluted the uncoupled magnon modes, we proceed to analyze their energies and magnetic field dependence with the aim of discussing the characteristic parameters governing spin ordering in the CoPS$_3$ antiferromagnet.

Since optical experiments specifically probe magnons with zero momentum ($k=0$), we can replace the full spin Hamiltonian given in Eq.~(\ref{H}) with its truncated two-sublattice representation:
 
\begin{eqnarray}
\hat{\cal H}_{2} & = & J_{\rm eff}\, {\bf S}_1 \cdot {\bf S}_2 + \hat{\cal H}_{a} - \mu_B
\sum_{i=1}^2 g_\alpha  B^\alpha S_i^\alpha \,
\label{H2subl}
\end{eqnarray}
For further details of such an approach see Ref.~\cite{Cho2023}. Here, $J_{\rm eff} = J_1+4J_2+3J_3$ is a net antiferromagnetic coupling between two opposite spins, $\hat{\cal H}_{a}$ is the biaxial magnetic anisotropy  (\ref{H}), and the last term is the Zeeman energy
 $\hat{\cal H}_Z$ written in the principal axes coordinate frame. Using the microscopic exchange parameters determined in \cite{Wildes2023}, we estimate
$J_{\rm eff} \approx 7.99$~meV. We use the Holstein-Primakoff representation for the $S=3/2$ spin operators to compute the $k=0$  modes described by 
$\hat{\cal H}_{2}$ in zero and finite magnetic fields.

In zero field ($B=0~T$), the two magnon gaps are expressed by
\begin{eqnarray}
\omega_{M_1} &= & 2S \sqrt{ (-2E)(J_{\rm eff}  + D - E) } \ ,
\nonumber  \\
\omega_{M_2} &= & 2S\sqrt{(D - E)(J_{\rm eff} - 2E)} \ ,
\label{H=0}
\end{eqnarray}
Using the microscopic constants from Ref.~\cite{Wildes2023}, we obtain $\omega_{M_1} = 115~\textrm{cm}^{-1}$ and
$\omega_{M_2} = 195~\textrm{cm}^{-1}$, which are larger than $109.5~\textrm{cm}^{-1}$ and 
$182~\textrm{cm}^{-1}$ determined experimentally in our work. In order to refine the microscopic constants, we use
the full magnetic field dependence of the bare magnon modes.

For a magnetic field perpendicular to the easy $x$ axis, spins form a canted antiferromagnetic structure. 
For $B\parallel z$, the magnon gaps $\omega_{M_1}$ and $\omega_{M_2}$ are expressed as
\begin{eqnarray}
\omega_{M_1} &= & 2S \cos\theta\sqrt{ (-2E)(J_{\rm eff}  + D - E) } \ ,
\label{eq:5}  \\
\omega_{M_2} &= & 2S\sqrt{(J_{\rm eff} - 2E)[J_{\rm eff} \sin^2\theta + (D - E) \cos^2\theta]} \ ,
\nonumber 
\end{eqnarray}
where the canting angle $\theta$ is given by
\begin{equation}
\sin\theta  = \frac{g_z \mu_B B}{2S(J_{\rm eff} + D - E)} \ .
\end{equation}
Similar expressions for $B\parallel y$ are
\begin{eqnarray}
\omega_{M_1} &= & 2S \sqrt{(J_{\rm eff}+ D - E)(J_{\rm eff} \sin^2\theta - 2E \cos^2\theta)} \ ,
\nonumber \\
\omega_{M_2} &= & 2S\cos\theta \sqrt{(D- E)(J_{\rm eff} - 2E)}  \ ,
\label{eq:4}   
\end{eqnarray}
with 
\begin{equation}
\sin\theta = \frac{g_y \mu_B B}{2S(J_{\rm eff} - 2E)} \ .
\end{equation}

Finally, for $B\parallel x$ configuration, spins preserve the up-down collinear structure till the spin-flop transition. Since the corresponding field is not reached in our experiments, we present only  the results for the collinear state: 
\begin{eqnarray} 
\omega_{M_{1,2}}^2  & = &  (J_{\rm eff} + D - 3E)^2S^2 - J_{\rm eff}^2 S^2 - (D + E)^2S^2 
\nonumber \\
&  + & (g_x \mu_B B)^2 \pm  2S\Bigl[
J_{\rm eff}^2(D + E)^2S^2
\label{Bx} \\
&+ & (g_x \mu_B B)^2 (D - 3E)(2J_{\rm eff} + D - 3E)\Bigr]^{1/2}\ .
\nonumber
\end{eqnarray}

\begin{table*}[]
\caption{Magnon-phonon coupling constants, bare magnon mode energies, and the magnetic interaction parameters of CoPS$_3$.}
\begin{tabular}{|c|c|c|c|c|c|l|l|l|l|l|l|l|l|l|}
\hline
~  & $\delta_1$~cm$^{-1}$  & $\delta_2$~cm$^{-1}$   & $\delta_3$~cm$^{-1}$   & $\omega_{P_1}$~meV  & $\omega_{P_2}$~meV  & $\omega_{P_3}$~meV  & $\omega_{M_1}$~meV    & $\omega_{M_2}$~meV  & $J_{\rm eff}$~meV    & $D$~meV   & $E$~meV    & g$_x$ & g$_y$ & g$_z$ \\ \hline
Our work & 2.5 & 10.8 & 12.5 & 14.8 & 19.6 & 19.5 & 13.7 & 22.8 & 9.9 & 4.3 & -0.7 & 2  & 2  & 4  \\ \hline
Ref.~\cite{Wildes2023} & - & - & - & - & - & - & 14.28 & 24.13 & 7.99 & 6.07 & -0.77 & -  & -  & -  \\ \hline
\end{tabular}
\label{Table1} \\
\end{table*}

\begin{figure*}[htp]
\centering
\includegraphics[width=16.8cm]{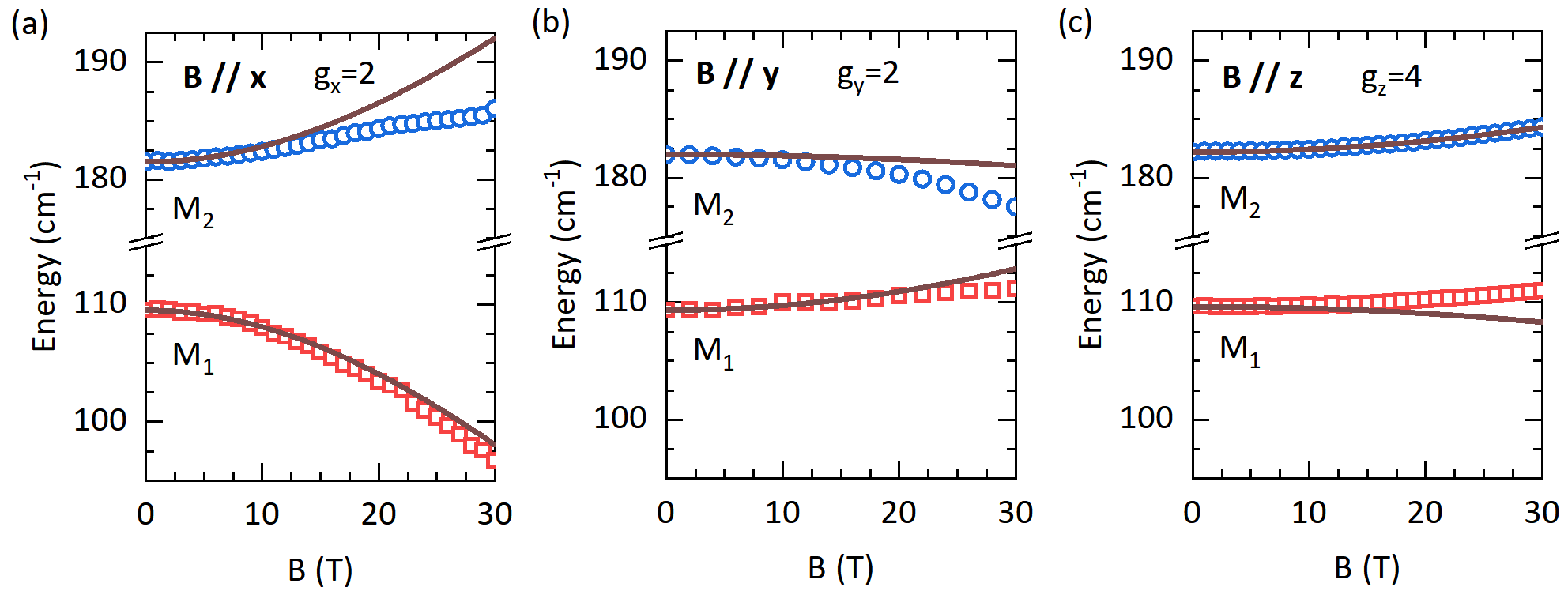}
\caption{(a)-(c) The theoretical evolution of magnon modes (M$_1$, M$_2$) as a function of magnetic field applied along magnetic $x$-, $y$-, and $z$-axes respectively (solid lines). The exchange and anisotropy parameters are considered to be $J_{eff}$ = 9.9~meV), $D$ = 4.3~meV and $E$ = -0.7~meV and the $g$-factors are: $g_x=2$, $g_y=2$, and $g_z=4$. The bare magnon mode energies obtained for the magnetic field configurations, presented in Fig.~\ref{fig:Fig3}a and ~\ref{fig:Fig4}a, are also shown in the respective plots (open symbols).}
\label{fig:Fig6}
\end{figure*}

Even though the microscopic parameters reported in Ref.~\cite{Wildes2023} do not reproduce the magnon gap energies, the trend of in-plane magnetic field dependence of $\omega_{M_1}$ and $\omega_{M_2}$ modes, as obtained using Eq.~(\ref{eq:4})-(\ref{Bx}) and shown in Fig.~S6 of SM~\cite{SuppInfo}, confirms that the in-plane magnetic field is predominantly oriented along the magnetic x-axis in Fig.~\ref{fig:Fig3} and along magnetic y-axis in Fig.~\ref{fig:Fig4}. To refine the microscopic parameters ($J_\text{eff}$, $D$, and $E$), we used Eq.~(\ref{H=0}) for the magnon gaps and the out-of-plane magnetic field dependence ($B \perp $~plane) of the M$_2$ mode (Eq.~(\ref{eq:5})), treating them as three constraints. Here we assume that the magnetic field remains perpendicular to the spin direction (even though it is canted by 10$^{\circ}$ along the out-of-plane direction). The resulting parameters, obtained by solving these equations (see Section~5 of SM~\cite{SuppInfo}), are listed in Table~\ref{Table1}. The $g$-factor along the magnetic $z$-axis is found to be 4 ($g_z = 4$). The theoretical field dependence of the magnon gaps, using this set of parameters, is presented by solid lines in Fig.~\ref{fig:Fig6}a-c, along with the experimental results shown by symbols. The theoretical curves reproduce the experimental trends reasonably well for in-plane configurations of the magnetic field when the $g$-factors are taken as $g_x = 2$ and $g_y = 2$. 

Notable deviations, particularly for the M$_1$ mode under $B \parallel z$-axis and at higher magnetic fields, may arise from several factors. First, even in the out-of-plane configuration ($B \parallel z$-axis), the field is not exactly orthogonal to the spin direction, introducing uncertainty into the parameter estimation. Similar misalignments are possible in the in-plane configurations, where the applied field may deviate from the exact magnetic axes. Furthermore, discrepancies may also originate from additional terms in the spin Hamiltonian that are not currently included in the theoretical model for CoPS$_3$. For example, the low monoclinic symmetry $C2/m$ at the Co site can lead to the following term in the crystal field Hamiltonian:
\begin{equation}
{\hat{\cal H}_a}^\prime = K \sum_i \bigl(S^x_i S^z_i  + S^z_i S^x_i \bigr) \ .
\end{equation}
Such a term causes Co spins to move out of the $ab$ plane at zero magnetic field. 
Such a tilt of about 10$^{\circ}$ has been experimentally observed in the neutron diffraction
study of CoPS$_3$ \cite{Wildes2017}. Computing the tilting angle by minimization of $\hat{\cal H}_a+{\hat{\cal H}_a}^\prime$
yields  $K\simeq 0.18(D+|E|)$, which is comparable to other anisotropy parameters. The magnetic field applied perpendicular to the plane is also not along the magnetic z-axis of CoPS$_3$. For magnetic fields up to 30~T, both the sublattice spins are expected to cant by about 10$^{\circ}$ for B~$\parallel z$-axis as can be obtained from Eq.~(\ref{eq:5}). However, in the presence of comparable zero-field tilting, the field-dependent canting is different for the two sublattice spins. A detailed numerical solution is thus required to reproduce experimentally the observed magnetic field dependence of the magnon modes, which is beyond the scope of this work.

While completing this manuscript, we became aware of a related study~\cite{mai2025spinlatt} that investigates spin–lattice entanglement in CoPS$_3$ via temperature-dependent Raman scattering.

\section{Conclusions}

In conclusion, the Raman scattering study reveals exceptionally strong and selective hybridization between magnons and phonons in the CoPS$_3$ antiferromagnet. A striking manifestation of this coupling is the pronounced splitting of two phonon-like modes observed at 152~cm$^{-1}$ and 158~cm$^{-1}$ in the antiferromagnetic phase, despite their nearly degenerate bare phonon origins. By modeling the extracted bare magnon modes and their magnetic field dependencies within existing theoretical frameworks, we propose an updated set of effective exchange and magnetic anisotropy parameters that refine our understanding of spin ordering and dynamics in CoPS$_3$. Our findings corroborate the strong anisotropy of the $g$-factor in CoPS$_3$.

\section{Acknowledgments}
Numerous valuable discussions with A. Wildes are acknowledged. The support of the LNCMI-CNRS, a member of the European Magnetic Field Laboratory (EMFL), is acknowledged.
M.P. acknowledges support from the European Union (ERC TERAPLASM No. 101053716) and the CENTERA2, FENG.02.01-IP.05-T004/23 project funded within the IRA program of the FNP Poland, co-financed by the EU FENG Programme.
R.S. acknowledges the financial support provided by the Ministry of Science and Technology in Taiwan under Project No. NSTC-113-2124-M-001-003 and No. NSTC-113-2112M001045-MY3, as well as support from Academia Sinica for the budget of AS-iMATE11412.
M.K. acknowledges support from the Ministry of Education (Singapore) through the Research Centre of Excellence program (grant EDUN C-33-18-279-V12, I-FIM), Academic Research Fund Tier 2 (MOE-T2EP50122-0012), and the Air Force Office of Scientific Research and the Office of Naval Research Global under award number FA8655-21-1-7026.

\providecommand{\noopsort}[1]{}\providecommand{\singleletter}[1]{#1}%

\newpage
\pagenumbering{gobble}

\begin{figure}[htp]
\includegraphics[page=1,trim = 18mm 18mm 18mm 18mm,
width=1.0\textwidth,height=1.0\textheight]{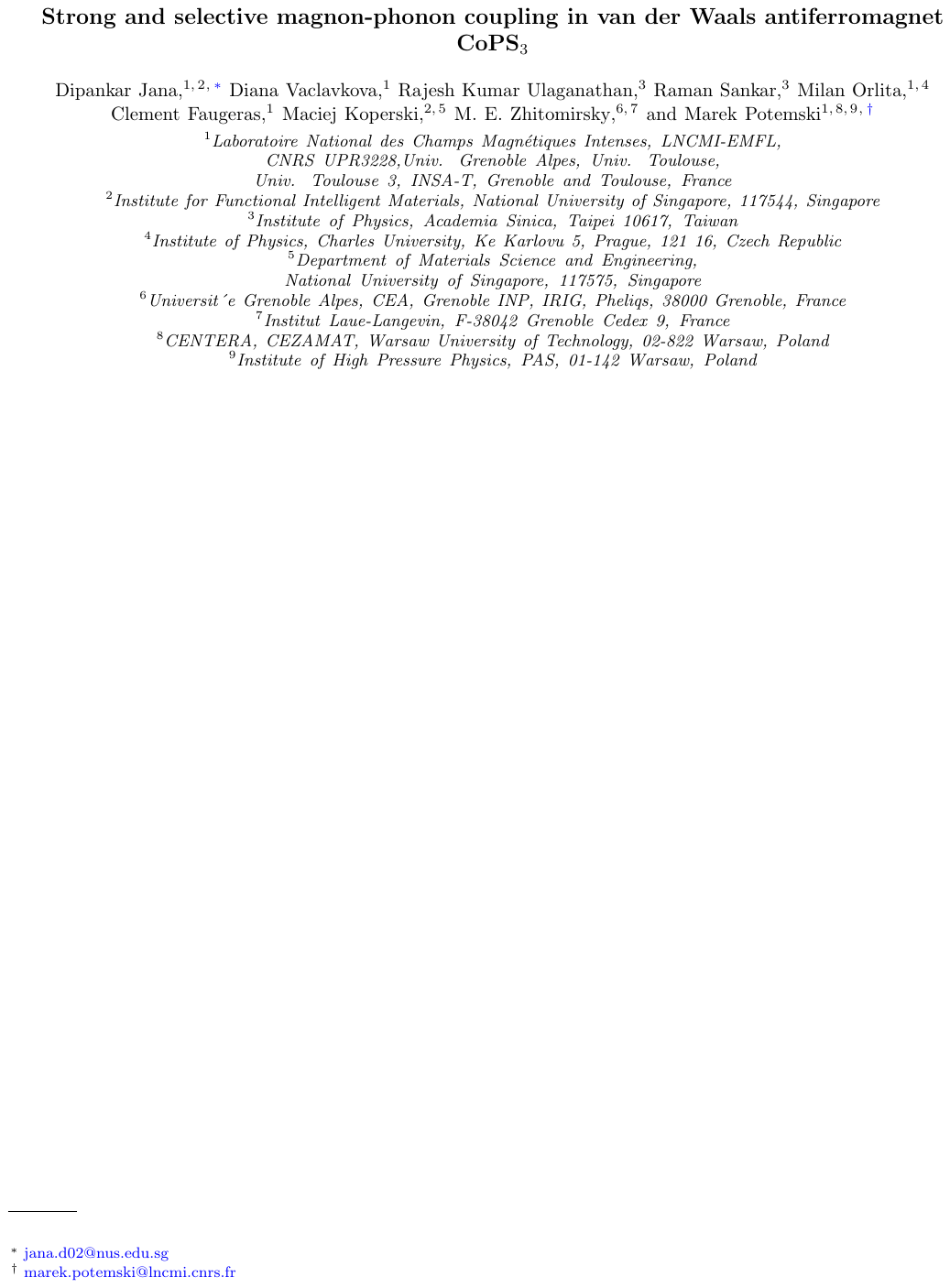}
\end{figure}

\newpage

\begin{figure}[htp]
   \includegraphics[page=2,trim = 18mm 18mm 18mm 18mm,
width=1.0\textwidth,height=1.0\textheight]{CoPS3_Magnon_SM.pdf}
\end{figure}
\newpage

\begin{figure}[htp]
   \includegraphics[page=3,trim = 18mm 18mm 18mm 18mm,
width=1.0\textwidth,height=1.0\textheight]{CoPS3_Magnon_SM.pdf}
\end{figure}

\begin{figure}[htp]
   \includegraphics[page=4,trim = 18mm 18mm 18mm 18mm,
width=1.0\textwidth,height=1.0\textheight]{CoPS3_Magnon_SM.pdf}
\end{figure}

\newpage

\begin{figure}[htp]
   \includegraphics[page=5,trim = 18mm 18mm 18mm 18mm,
width=1.0\textwidth,height=1.0\textheight]{CoPS3_Magnon_SM.pdf}
\end{figure}
\newpage

\begin{figure}[htp]
   \includegraphics[page=6,trim = 18mm 18mm 18mm 18mm,
width=1.0\textwidth,height=1.0\textheight]{CoPS3_Magnon_SM.pdf}
\end{figure}

\begin{figure}[htp]
   \includegraphics[page=7,trim = 18mm 18mm 18mm 18mm,
width=1.0\textwidth,height=1.0\textheight]{CoPS3_Magnon_SM.pdf}
\end{figure}

\begin{figure}[htp]
   \includegraphics[page=8,trim = 18mm 18mm 18mm 18mm,
width=1.0\textwidth,height=1.0\textheight]{CoPS3_Magnon_SM.pdf}
\end{figure}

\end{document}